\documentclass[sn-mathphys]{sn-jnl}
\usepackage{bm}
\usepackage{amsmath}

\jyear{2022}%
\theoremstyle{thmstyleone}%
\theoremstyle{thmstyletwo}%

\theoremstyle{thmstylethree}%

\begin{document}

\title[$py$GWBSE: A high throughput workflow package for GW-BSE calculations]{$py$GWBSE: A high throughput workflow package for GW-BSE calculations}

\author[1]{\fnm{Tathagata} \sur{Biswas}}\email{tbiswas3@asu.edu}

\author*[1]{\fnm{Arunima K.} \sur{Singh}}\email{arunimasingh@asu.edu}

\affil*[1]{\orgdiv{Department of Physics}, \orgname{Arizona State University}, \orgaddress{\city{Tempe}, \postcode{85287}, \state{Arizona}, \country{USA}}}

\abstract{ We develop an open-source python workflow package, \emph{py}GWBSE
to perform automated first-principles calculations within the GW-BSE (Bethe-Salpeter) framework. GW-BSE is a many body perturbation theory based approach to explore the
quasiparticle (QP) and excitonic properties of materials.  The GW approximation
has proven to be effective in accurately predicting bandgaps of a wide range of
materials by overcoming the bandgap underestimation issues of the more widely used
density functional theory (DFT). The BSE formalism, in spite of being
computationally expensive, produces absorption spectra directly comparable with
experimental observations. The \emph{py}GWBSE package achieves complete
automation of the entire multi-step GW-BSE computation, including the
convergence tests of several parameters that are crucial for the accuracy of
these calculations. \emph{py}GWBSE is integrated with \emph{Wannier90}, a
program for calculating maximally-localized wannier functions, allowing the
generation of QP bandstructures. \emph{py}GWBSE also enables the automated creation
of databases of metadata and data, including QP and excitonic properties, which
can be extremely useful for future material discovery studies in the field of
ultra-wide bandgap semiconductors, electronics, photovoltaics, and
photocatalysis.} 

\maketitle
\section{Introduction}
\label{intro}

Obtaining materials with properties that are optimized for a particular
application traditionally relies on time-consuming and expensive experimental
work. However, in recent years an alternative paradigm in the field of material
discovery has emerged through the availability of modern massive supercomputing
resources, development of first-principles methodologies, and ingenious
computational algorithms. These advancements have pushed the boundaries of
materials simulations, making them faster, more cost-effective, efficient and
accurate. Consequently, high-throughput materials simulations have emerged as
a tool for creating large databases and  screening materials from these
databases to identify candidate materials for applications in photocatalysis
\cite{singh2019robust, wu2013first}, energy storage \cite{kirklin2013high,
hautier2011phosphates}, piezoelectrics \cite{choudhary2020high}, 
electrocatalysis \cite{greeley2006computational} etc. 

However, applying similar approaches to applications related to optical and
transport properties of materials have been hindered by a few technical
obstacles. Density functional theory (DFT), the most widely used tool in
computational high-throughput materials discovery studies are designed to
explore ground state properties of a system and has been remarkably successful
in predicting structural, mechanical, electronic, and thermal properties
\cite{jones2015density}. However, studying the excited state properties of a system, such as optoelectronic or transport properties using DFT requires the
interpretation of Kohn-Sham (KS) eigenvalues as energies involved in adding an
electron to a many-electron system or subtracting one from it (QP energies).
Following such a procedure, one often encounters the infamous bandgap
underestimation problem due to the derivative discontinuity of the
exchange-correlation energy \cite{perdew1985density}. 

A more rigorous approach to computing QP energies and accurate bandgap is
applying many body perturbation theory (MBPT) within the GW approximation
\cite{hedin1965new}. Using this formalism, one computes the QP energies by
calculating the first-order perturbative correction to the KS eigenvalues by
approximating self-energy as a product of one particle Green's function (G)
and screened Coulomb interaction (W). It has been shown that MBPT within the
GW approximation is particularly useful in computing QP properties of a wide
variety of semiconductors and insulators \cite{onida2002electronic} without
requiring  any ad-hoc introduction of mixing parameters like those needed for
hybrid functionals used in DFT \cite{muscat2001prediction,
vines2017systematic}. 

Additionally, the study of QP properties using GW formalism enables us to
compute several transport properties of materials that are inaccessible from a
DFT calculation.  For example, QP lifetimes calculated from GW calculations can
be directly used to estimate impact ionization rates, a very useful parameter
in the study of high-field transport of wide bandgap materials
\cite{kotani2010impact}. In case of low field transport, in addition to the
obvious importance \cite{darancet2007ab} of including GW corrections to KS
eigenvalues and carrier mobilities, it has been shown that one needs to include
the effects of GW correction on the orbital character of the relevant KS
wavefunctions to obtain accurate transport properties of molecular junctions
\cite{rangel2011transport}. 

The optical and transport properties of semiconductors to a large extent are
defined by the presence of intentional dopants or unintentional defects. GW
formalism has emerged as a powerful approach that complements experiments and
has become reliable enough to serve as a predictive tool for crucial point
defect properties such as charge transition levels and F-center
photoluminescence spectra in semiconductors
\cite{biswas2019electronic,freysoldt2014first}. Moreover, calculations based on
MBPT using GW approximation has been successfully applied to estimate
non-radiative recombinations such as Auger recombination rates
\cite{kioupakis2011indirect,mcallister2015auger} which are very useful for
optical applications. Auger recombination mechanism has been shown to cause
significant efficiency loss in InGaN-based light-emitting diodes (LEDs), when
operating at high injected carrier densities.

The necessity of the BSE methodology lies in the fact that even after including
GW corrections the optical spectrum calculated within the independent-particle
picture shows significant deviations from experimental results, as not only
the absorption energies can be wrong, but often the oscillator strength of the
peaks can deviate from the experiment by a factor of 2 or more. Moreover, it can
not describe bound exciton states, which are particularly important in 
systems of reduced dimensions \cite{ugeda2014giant}. The reason
being, independent particle picture can't include electron-hole interactions
(excitonic effects) which requires an effective two-body approach
\cite{rohlfing2000electron}. This can be achieved by evaluating the two-body
Green’s function G$_2$  and formulating an equation of motion for G$_2$, known
as the Bethe-Salpeter equation (BSE) \cite{rohlfing2000electron}.

Despite its obvious indispensability, high-throughput computational material
discovery studies for light-matter interaction related applications so far have
not been able to incorporate the QP or excitonic properties of materials using
GW-BSE formalism. Two main challenges for such an
endeavor is the efficient convergence of multiple parameters and the tractability of the
huge computational cost associated with the multi-step GW-BSE formalism. GW-BSE
calculations are extremely sensitive to multiple interdependent convergence
parameters such as the number of bands included in the GW self-energy
calculation or the number of $k$-points used to sample the Brillouin zone (BZ)
in the BSE calculation etc. 

In this article, we introduce the open-source Python package, $py$GWBSE, which automates the entire GW-BSE calculation using first-principles simulations software Vienna Ab-initio Software Package
($VASP$) \cite{hafner2008ab}. This package enables automated input file generation, submission to supercomputing platforms, analysis of post-simulation
data, and storage of metadata and data in a MongoDB database. Moreover,
pyGWBSE is capable of handling multiple convergence parameters associated
with the GW-BSE formalism. Using this package, high-throughput computation of various electronic and optical properties is possible in a systematic and efficient manner. For example, the QP energies, bandstructures, and density of states can be computed using both the one-shot G$_0$W$_0$ and partially self-consistent GW$_0$ level of the GW formalism. The package enables automated BSE computations yielding the real and imaginary part of the dielectric function (incorporating electron-hole interaction), the exciton energies, and their corresponding oscillator strengths. DFT bandstructures, orbital resolved density of states (DOS), electron/hole effective masses, band-edges, real and imaginary parts (absorption spectra) of the dielectric function, and static dielectric tensors can also be computed using $py$GWBSE.

The package is being continuously developed and the latest version can be
obtained from the GitHub repository at ${\it
https://github.com/cmdlab/pyGWBSE}$. $py$GWBSE  is built upon existing open source Python packages such as, $pymatgen$ \cite{ong2013python}, $Fireworks$ \cite{jain2015Fireworks}, and $\it atomate$
\cite{mathew2017atomate}. To obtain the QP bandstructure we use the
$Wannier90$ \cite{mostofi2008wannier90}, a program for calculating
maximally-localized Wannier functions to perform the interpolation required to
obtain QP bandstructure with reduced computational cost.

$py$GWBSE enables high-throughput
simulations of highly reliable and efficient $ab$-$initio$ approaches thus enabling future materials screening studies, the creation of large databases of high-quality computed properties of materials, and in turn machine learning model development. Hence, $py$GWBSE could serve as a catapult to the next generation of technological advances in the field of power electronics, optoelectronics, photovoltaics, photocatalysis, etc.    

In the following sections, we present an overview of the underlying methodology, describe the workflow architecture, discuss the algorithms that were developed to perform the multi-step convergences, and benchmark the results obtained from the $py$GWBSE workflow against experimental data in the literature.

\section{$py$GWBSE: Underlying Methodology and Codes}
\label{methods}

The first principles calculations in the $py$GWBSE package are performed using
one of the most well-known packages, $VASP$. $VASP$ is a first principles
computer program for atomic scale materials modeling. It is capable of using
the projector-augmented wave (PAW) method \cite{blochl1994projector} and it
comes with a rigorously tested pseudopotential library. $VASP$ provides the
accuracy of the full-potential linearized augmented-plane-wave (FLAPW)
\cite{singh1994planes} method, but is computationally less expensive than most
of the traditional plane wave-based methods \cite{hafner2008ab}. Currently, in
addition to DFT, GW, and BSE, it supports various other post-DFT methods such
as TD-DFT, ACFDT, 2nd-order Møller-Plesset perturbation theory and is under
constant development, which opens up the possibility of implementing these
methodologies in the $py$GWBSE workflow in the future. Additionally, $VASP$ is
very efficiently parallelized and can utilize the potential of modern computers
of both CPU and GPU-based architectures.  

The $py$GWBSE package is capable of computing several material properties using
the DFT, GW, and BSE methodologies. A brief summary of GW-BSE methodology has
been presented in section \ref{GW methods} (GW) and \ref{BSE methods} (BSE).
Using $py$GWBSE, properties such as bandstructures, the orbital resolved density
of states (DOS), electron/hole effective masses, band-edges, real and imaginary
part (absorption spectra) of the dielectric function ($\epsilon(\omega)$), and
static dielectric tensors can be computed using the DFT methodology. In the $py$GWBSE package, the GW formalism can be used to compute QP energies both at one-shot G$_0$W$_0$ and partially
self-consistent GW$_0$ level of accuracy. The package uses maximally localized
wannier functions (MLWF) to compute the electronic structure at the QP level of
accuracy but at a significantly reduced computation cost. Using this package, the
BSE methodology can be used to compute the real and imaginary parts of the
dielectric function (incorporating electron-hole interaction), the exciton
energies, and their corresponding oscillator strengths. 

\subsection{Theoretical background and parameter convergence in $py$GWBSE}
\label{theory}

\begin{figure}
\centering
\includegraphics[width=0.8\linewidth]{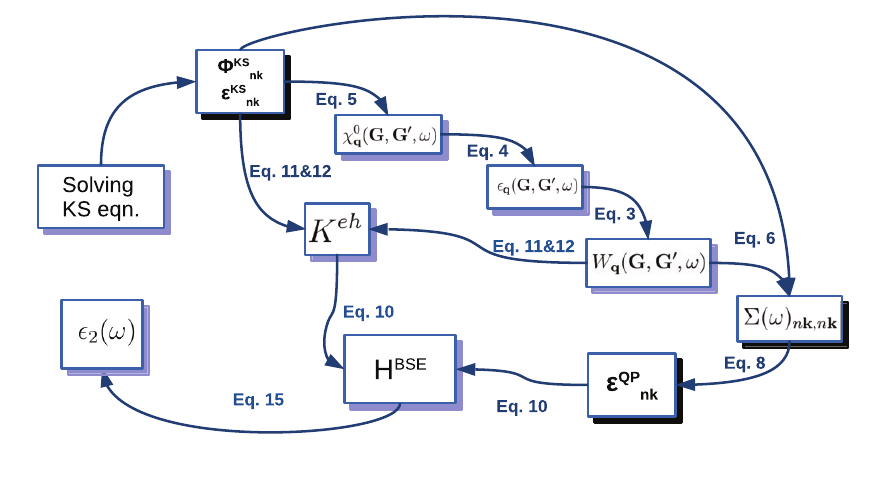}
\caption{A schematic summarizing the GW-BSE framework. All the equations as well as the key quantities mentioned in section~\ref{GW methods} and section~\ref{BSE methods}  are drawn to provide a condensed visual representation.} 
\label{fig1}
\end{figure}

In the following two sections, we describe the $py$GWBSE package's
computational framework with a particular emphasis on all the crucial
computational parameters which are needed to be converged for obtaining
accurate results. Fig. \ref{fig1} provides a condensed diagrammatic representation of the GW-BSE framework described in the following two subsections; showing the interconnection between the key equations and the various physical quantities. Detailed discussions about the GW-BSE methodology can be found in several review articles \cite{onida2002electronic, leng2016gw, faber2014excited, blase2018bethe}. 

Additionally, we also present the rationale
behind the strategies that we adopt to reduce the computational cost of the
convergence calculations and thus make the computations more time-efficient. A discussion of the convergence parameters for the DFT ground state
calculations can be found in the existing literature
\cite{jain2011high,kresse2018j}. In these studies, plane wave energy cutoff was set to 1.3 times the maximum energy cutoff specified in the pseudopotentials and $k$-grid
 set as (500)/n points, where n represents the number of atoms in the unit cell
distributed as uniformly as possible in $k$-space. It resulted in total energy convergence of 15 meV/atom for 96$\%$ of 182 chemically diverse materials
\cite{jain2011high}. In the following, sections \ref{GW methods} and \ref{BSE methods}, we discuss that the same choice for the plane wave energy cutoff but a higher $k$-grid density is required to converge the 
GW-BSE calculations. Additionally, other convergence
parameters are needed for the GW-BSE calculations, and the strategies to converge them are elaborated through examples.

\subsubsection{Obtaining converged QP energies using GW methodology}
\label{GW methods}

The QP energies are the energies for adding an electron to a many-electron
system or subtracting one from it. Within the MBPT they can be calculated by
solving the following equation \cite{hedin1965new},

\begin{equation}
	(T+V_{n-e}+V_{H}-E_{n{\bf k}}^{QP})\psi^{QP}_{n{\bf k}}({\bf r})+\int d^3{\bf r'} 
	\Sigma ({\bf r},{\bf r'}, E_{n{\bf k}}^{QP}) \psi^{QP}_{n{\bf k}}({\bf r'}) = 0
\label{g1}
\end{equation}

where, $T$ is the kinetic energy operator, $V_{n-e}$ is the operator to account
for the nuclear(ion)-electronic interaction, $V_H$ is the Hartree potential,
$\Sigma$ is the self-energy operator and $\bf r$ is the position vector 
of the electron. $E_{n{\bf k}}^{QP}$ and $\psi^{QP}_{n{\bf k}}$s are the 
QP energies and wavefunctions for $n^{th}$ band with wavevector $\bf k$.

GW approximation provides a practical route to compute the self energy operator, $\Sigma ({\bf r},{\bf r'}, \omega)$, from KS wavefunctions, $\psi_{n{\bf k}}$, 
and energies, $\epsilon_{n{\bf k}}$, through
one particle Green's function, $G({\bf r}, {\bf r'}, \omega)$.\cite{hedin1965new,hybertsen1986electron,shishkin2006implementation} Within the GW
approximation $\Sigma$ can be written as,

\begin{equation}
	\Sigma ({\bf r},{\bf r'}, \omega)= \frac{i}{4\pi} \int_{-\infty}^{\infty}
	e^{i\omega'\delta} G({\bf r}, {\bf r'}, \omega+\omega')W({\bf r},{\bf r'}, \omega') d\omega'
\label{g2}
\end{equation}

where, $W$ is the screened Coulomb interaction, $\omega$ is the frequency and
$\delta$ is the positive infinitesimal.

$W$ can be computed from $\psi_{n{\bf k}}$ and $\epsilon_{n{\bf k}}$ through
independent particle polarizability and frequency dependent dielectric matrix
utilizing the following three equations and taking advantage of the Random
Phase Approximation (RPA) \cite{hybertsen1987ab}.

\begin{equation}
	W_{\bf q}({\bf G}, {\bf G'},\omega)=4 \pi e^2 \frac {1}{\lvert {\bf q}
	+ {\bf G} \rvert} \epsilon^{-1}_{\bf q} ({\bf G}, {\bf G'},\omega)
	\frac {1}{\lvert {\bf q}+{\bf G'} \rvert}
\label{g3}
\end{equation}

The dielectric matrix, $\epsilon_{\bf q}({\bf G}, {\bf G'},\omega)$ is related
to $\chi$ as \cite{hybertsen1987ab},

\begin{equation}
	\epsilon_{\bf q}({\bf G}, {\bf G'},\omega)=\delta_{{\bf G}{\bf G'}} 
	- \frac {4 \pi e^2}{\lvert {\bf q}+{\bf G}\rvert \lvert{\bf q}+{\bf G'} \rvert} 
	\chi^0_{\bf q}({\bf G}, {\bf G'},\omega)
\label{g4}
\end{equation}

Within RPA, independent particle polarizibility, $\chi^0_{\bf q}({\bf G}, {\bf
G'},\omega)$, is calculated as \cite{hybertsen1987ab},

\begin{multline}
	\chi^0_{\bf q}({\bf G}, {\bf G'},\omega) =\Omega^{-1} \Sigma_{n,n',{\bf
	k}} 2 w_{\bf k} (f_{n,{{\bf k}-{\bf q}}} - f_{n,{\bf k}}) \times \\ \frac 
	{\langle \psi_{n' {\bf k}-{\bf q}} \lvert e^{-i({\bf q}+{\bf G}){\bf
	r}} \rvert \psi_{n {\bf k}}\rangle \langle \psi_{n{\bf k}} \lvert
	e^{i({\bf q}+{\bf G'}){\bf r'}} \lvert \psi_{n' {\bf k}-{\bf
	q}}\rangle} {\omega+\epsilon_{n'{\bf k}-{\bf q}}-\epsilon_{n{\bf k}}+ i
	\eta sgn[\epsilon_{n'{\bf k}}-\epsilon_{n{\bf k}}]}
\label{g5}
\end{multline}

where, $w_{\bf k}$ is the $k$-point weight, the $f_{n,{\bf k}}$ are the one
electron occupancy of the corresponding states, ${\bf q}$ is the Bloch wave
vector, $\bf G$ is the reciprocal lattice vector and $\eta$ is the
infinitesimal complex shift.

Once the screened Coulomb interaction, $W$, is computed, the diagonal matrix
elements of self-energy operator, $\Sigma(\omega)_{n{\bf k},n{\bf k}}$, can be
obtained using \cite{hybertsen1986electron},

\begin{multline}
	\Sigma(\omega)_{n{\bf k},n{\bf k}}=\Omega^{-1} 
	\Sigma_{{\bf q}{\bf G}{\bf G'}} \Sigma_{n'}
	\frac{i}{2\pi} \int^{\infty}_0 d{\omega'} 
	W_{\bf q} ({\bf G},{\bf G'},\omega') \\
	\times 
	\langle \psi_{n {\bf k}} \lvert e^{i({\bf q}+{\bf G}){\bf r}} 
	\rvert \psi_{n' {\bf k}-{\bf q}} \rangle
	\langle \psi_{n' {\bf k}-{\bf q}} \lvert e^{-i({\bf q}+{\bf G'}){\bf r'}} 
	\rvert \psi_{n {\bf k}}\rangle \\
	\times 
	(\frac {1} {\omega+\omega'-\epsilon_{n' {\bf k}-{\bf q}} 
	+ i\eta sgn(\epsilon_{n' {\bf k}-{\bf q}}-\mu)} 
	+\frac {1} {\omega-\omega'-\epsilon_{n' {\bf k}-{\bf q}} 
	+ i\eta sgn(\epsilon_{n' {\bf k}-{\bf q}}-\mu)})
\label{g6}
\end{multline}

where, $\mu$ is the Fermi energy. In the non-self-consistent GW calculation
(also known as G$_0$W$_0$) G$_0$ and W$_0$ are calculated using KS eigenvalues and
eigenfunctions. The wavefunction of QP Hamiltonian (Eqn. \ref{g1}) is
approximated as the DFT wavefunction and the QP energies are computed to first
order as,

\begin{equation}
	E^{QP}_{n{\bf k}} = Re[\langle \psi_{n{\bf k}} 
	\lvert T+V_{n-e}+V_H+\Sigma(E^{QP}_{n{\bf k}}) \rvert \psi_{n{\bf k}} \rangle]
\label{g7}
\end{equation}

Since Eqn. \ref{g7} requires the values of $E^{QP}_{n{\bf k}}$, the equation
must be solved by iteration.  Using the usual Newton-Raphson method for root
finding, one can obtain the following update equation
\cite{hybertsen1986electron},

\begin{equation}
	E^{QP}_{n{\bf k}} \leftarrow E^{QP}_{n{\bf k}} + Z_{n {\bf k}} Re
	[\langle \psi_{n{\bf k}} \lvert T+V_{n-e}+V_H+\Sigma(E^{QP}_{n{\bf k}}) \rvert
	\psi_{n{\bf k}} \rangle -  E^{QP}_{n{\bf k}}] 
\label{g8}
\end{equation}

where $Z_{n {\bf k}}$ is the renormalization factor and can be calculated as,

\begin{equation}
	Z_{n{\bf k}} = (1-Re(\langle \psi_{n{\bf k}} 
	\lvert \frac {\partial \Sigma (\omega)}{\partial \omega} \rvert 
	\psi_{n{\bf k}} \rangle))^{-1}
\label{g9}
\end{equation}

The iteration starts from DFT eigenvalues and if one stops after the first
iteration the QP energies are obtained within G$_0$W$_0$ approximation. One can
also continue to obtain QP energies which are self-consistently converged.
This scenario is referred to as self-consistent GW approximation (scGW)
\cite{shishkin2007self}. 

In the $py$GWBSE workflow, both the G$_0$W$_0$ and scGW are implemented. In the partially self-consistent scGW approximation the $G$ is updated self-consistently until convergence is reached but the $W$ is kept unchanged, thus the scGW is also referred to as GW$_0$. A full update of the $G$ and  $W$ is seldom adopted and thus is not included in the $py$GWBSE package. In fact, it has been shown by several
studies \cite{delaney2004comment,schone1998self,tiago2004effect} pertaining to
free-electron gas, metals, and semiconductors that fully self-consistent GW
calculations, without vertex corrections, lead to an overestimation of
bandgaps. 

There are three crucial parameters in a GW calculation that need to be
converged to obtain accurate results, namely- 

\begin{itemize}
\item{Number of plane waves used to expand the screened Coulomb operator, 
    $W_q({\bf G},{\bf G'},\omega)$. This parameter can be specified by
using {\it ENCUTGW} in the $VASP$ implementation.} 
\item{Number of frequency grid
	points used in Eqn. \ref{g6} for the frequency integration. This parameter can
		be specified by using {\it NOMEGA} in the $VASP$ implementation.}
\item{Number of bands used in Eqn. \ref{g5} and \ref{g6} for the summation. This parameter can
	be specified by using {\it NBANDS} in the $VASP$ implementation.}
\end{itemize}

\begin{table}[h]
\begin{center}
\begin{minipage}{174pt}
\caption{QP gaps of 9 materials computed using a screened coulomb cutoff, $ENCUTGW$, of 100 eV (E$^{100}_g$), 150 eV (E$^{150}_g$) and 200 eV
(E$^{200}_g$). The table shows that 200 eV cutoff is sufficient to converge QP
gaps within 0.1 eV for all these materials.} 
\label{tab1}%
\begin{tabular}{@{}lllll@{}}
\toprule
mp-id & Formula & E$^{100}_g$ & E$^{150}_g$ & E$^{200}_g$ \\
\midrule
        mp-149  &  Si  &   1.18  &   1.20  &   1.20 \\
        mp-390  &  TiO$_2$  &   3.89  &   3.89  &   3.89 \\
        mp-66  &  C  &   5.24  &   5.40  &   5.46 \\
        mp-2133  &  ZnO  &   2.66  &   2.68  &   2.67 \\
        mp-804  &  GaN  &   2.84  &   2.87  &   2.86 \\
        mp-2624  &  AlSb  &   1.69  &   1.69  &   1.70 \\
        mp-1434  &  MoS$_2$  &   2.33  &   2.34  &   2.34 \\
        mp-984  &  BN  &   5.45  &   5.46  &   5.46 \\
        mp-22862  &  NaCl  &   7.71  &   7.72  &   7.73 \\
\botrule
\end{tabular}
\end{minipage}
\end{center}
\end{table}

\begin{table}[h]
\begin{center}
\begin{minipage}{174pt}
\caption{QP gap of 9 materials computed using {\it NOMEGA}  of 50 (E$^{50}_g$), 65
(E$^{65}_g$) and 80 (E$^{80}_g$). {\it NOMEGA} refers to the number of frequency grid points used in the numerical integration for evaluating the GW self-energy. The table shows
that 80 frequency grid points are sufficient to converge QP gaps within 0.1 eV
for all of these materials.}
\label{tab2}%
\begin{tabular}{@{}lllll@{}}
\toprule
    mp-id & Formula & E$^{50}_g$ & E$^{65}_g$ & E$^{80}_g$ \\
\midrule
    mp-149  &  Si  &   1.18  &   1.18  &   1.18 \\
    mp-390  &  TiO$_2$  &   3.89  &   3.80  &   3.75 \\
    mp-66  &  C  &   5.24  &   5.24  &   5.23 \\
    mp-2133  &  ZnO  &   2.66  &   2.48  &   2.38 \\
    mp-804  &  GaN  &   2.84  &   2.84  &   2.84 \\
    mp-2624  &  AlSb  &   1.69  &   1.67  &   1.66 \\
    mp-1434  &  MoS$_2$  &   2.33  &   2.28  &   2.25 \\
    mp-984  &  BN  &   5.45  &   5.44  &   5.43 \\
    mp-22862  &  NaCl  &   7.71  &   7.68  &   7.67 \\
\botrule
\end{tabular}
\end{minipage}
\end{center}
\end{table}

In principle, the QP energies are to be converged w.r.t all three of the
parameters mentioned above. Therefore, in our workflow, we have implemented the
convergence tests for QP energies w.r.t. these parameters namely,  {\it
ENCUTGW}, {\it NOMEGA}, and {\it NBANDS}. 

Table \ref{tab1} and \ref{tab2} show the QP gaps of 9 materials computed with different values of {\it ENCUTGW} and {\it NOMEGA}. Table \ref{tab1} shows the QP gaps computed with {\it ENCUTGW} values of 100, 150 and 200 eV, whereas Table \ref{tab2} shows the QP gaps computed with {\it NOMGEA} values of 50, 65, and 80. From Table \ref{tab1} we can see that, a {\it ENCUTGW} value of 150 eV is sufficient for obtaining a QP gap value that is converged
within $\sim$ 0.1 eV for almost all the materials except diamond (C), for which
we need an {\it ENCUTGW} value of 200 eV. Table \ref{tab2} shows that we need
a {\it NOMEGA} of 80 to converge the QP gap within 0.1 eV for all the selected
materials. Thus, as recommended by the $VASP$ manual, $ENCUTGW$ value of 2/3 $\times$ $ENCUT$ and $NOMEGA$ value of 50--100, would be sufficient to obtain accurate QP energies for
a wide range of materials. Furthermore, based on the convergence of the 9 materials in Tables \ref{tab1} and \ref{tab2} we surmise that given the variety in the chemical compositions and crystal structures of these materials, it is likely that a value of 200 eV for {\it ENCUTGW} and 80 for {\it NOMEGA} may be sufficient to converge QP gaps of a variety of other materials within 0.1 eV. 

The third parameter, ${\it NBANDS}$, however, needs to be converged for every material.
Despite many efforts to eliminate or reduce the need of including large number
of empty orbitals in the GW calculation, it still is one of the major
computation costs of a GW calculation. While several methods have been proposed
to reduce the total number of empty orbitals in a GW calculation such as
replacing actual KS orbitals with approximate orbitals generated using a reduced
basis set, truncation of the sum over empty orbitals to a reduced number, and
adding the contribution of the remaining orbitals within the static (COHSEX)
approximation and modified static remainder approach
\cite{deslippe2013coulomb}. These methods are not currently implemented in the
$VASP$ package. However, in the future, if these methods are implemented, they can
be easily incorporated into the $py$GWBSE package and would reduce the
computational costs of GW-BSE calculations significantly. Meanwhile, we
strongly suggest performing convergence tests w.r.t {\it NBANDS} using $py$GWBSE to obtain accurate results.

\subsubsection{Obtaining converged absorption spectra by solving BSE}
\label{BSE methods}

Studying optical electron-hole excitations is an effective two-body problem. In
most cases single-particle picture of individual quasi-electron and quasi-hole
excitations are not enough. We need to include electron-hole interactions as
well. We can work with two-body Green’s function $G_2$ on the basis of the
one-body Green’s function $G_1$, which can be described by the GW
approximation. We can use QP electron and hole states of $G_1$ and their QP
energies to estimate the electron-hole interactions. The equation of motion for
$G_2$ is known as the Bethe-Salpeter equation \cite{strinati1988application}
and is very useful in the study of correlated electron-hole excitation states
also known as excitons. 

Following Strinati \cite{strinati1988application}, Rohlfing and Louie
\cite{rohlfing2000electron}, the BSE can be written as a generalized
eigenvalue problem and the electron-hole excitation states 
can be calculated through the solution of BSE. For each exciton state 
$S$, within Tamm-Dancoff approximation the BSE can be written as,

\begin{equation}
	(\epsilon^{QP}_{c} - \epsilon^{QP}_{v}) A_{vc{\bf k}} + \Sigma_{v'c'{\bf k'}} 
	\langle {{vc{\bf k}}} \lvert K^{eh} \rvert {v'c'{\bf k'}}\rangle = \Omega^S A_{vc{\bf k}}  
\label{b1}
\end{equation}

where, $A_{vc{\bf k}}$ is the exciton wavefunction, $\Omega^S$ is the
excitation energy, $\epsilon^{QP}_{c {\bf k}}$ and $\epsilon^{QP}_{v {\bf k}}$
are the QP energies of the conduction ($\lvert c{\bf k}\rangle$) and valence states
($\lvert v{\bf k}\rangle$) which is computed using the GW methodology  discussed in
the previous section. The electron-hole interaction kernel, $K^{eh}$, can be
separated in two terms, $K^{eh}$ = $K^d$ + $K^x$, where $K^d$ is the screened
direct interaction term and $K^x$ is the bare exchange interaction term. Within
the GW approximation for $\Sigma$, in the basis of the single-particle orbitals
in real space ($\phi_{c/v}(\bf x)$, the KS orbitals obtained from DFT calculations), they are defined in the following way,

\begin{equation}
	\langle vc \lvert K^x \rvert v'c' \rangle = \int d{\bf x} d{\bf x'} \phi^*_{c}({\bf x})
	\phi_{v}({\bf x}) v({\bf r},{\bf r'}) \phi^*_{v'}({\bf x'})
	\phi_{c'}({\bf x'}) 
\label{b2}
\end{equation}

\begin{equation}
	\langle vc \lvert K^d \rvert v'c' \rangle = - \int d{\bf x} d{\bf x'}
	\phi^*_{c}({\bf x})  \phi_{c'}({\bf x}) W({\bf r},{\bf r'}; \omega=0)
	\phi^*_{v'}({\bf x'}) \phi_{v}({\bf x'}) 
\label{b3}
\end{equation}

Once we have the solutions of the BSE Hamiltonian, we can construct
$\epsilon_2(\omega)$ which incorporates excitonic effects from the solutions of
the modified BSE, 

\begin{multline}
	\epsilon_2(\omega)  = \frac {8 \pi^2 e^2} {\omega^2} 
	\sum_s \lvert \hat{\lambda} \cdot \langle 0 \lvert  {\bf v} \rvert S 
	\rangle \rvert ^2 \delta (\omega - \Omega^S) \\ 
         = \frac {8 \pi^2 e^2} {\omega^2} \sum_s \lvert 
	 \sum_{vc{\bf k}} A^S_{vc{\bf k}} \hat{\lambda} \cdot
	 \langle v{\bf k} \lvert {\bf v} \lvert c{\bf k} 
	 \rangle \rvert ^2 \delta (\omega - \Omega^S)
\label{b4}
\end{multline}

where $\hat{\lambda}$ is the polarization vector, and ${\bf v}$ is the velocity
operator along the direction of the polarization of light, $\hat{\lambda}$.
The real part of the dielectric function $\epsilon_1(\omega)$, can be obtained
by integration of $\epsilon_2(\omega)$ over all frequencies via Kramers–Kronig
relations. 

Converging the BSE absorption spectra with the number of $k$-points used to
sample the BZ is one of the most computationally demanding tasks in a GW-BSE
calculation. In this study, we propose a strategy to achieve this convergence
with a significant reduction in the computational cost involved. We
propose to obtain a convergent $\epsilon_2(\omega)$ within an independent particle
picture (RPA) and use the same $k$-mesh to perform the BSE calculation. This
strategy is expected to be useful because of the following reasons.  

There are two aspects of the convergence of $\epsilon_2(\omega)$ w.r.t
$\rho_k$. Firstly, due to band dispersion, one needs to include all the
occupied-unoccupied transitions throughout the BZ to obtain absorption spectra
that are converged. As a result, one needs to use a very dense $k$-grid for
materials that have a stronger dispersion of bands near the gap. Secondly, the
electron-hole interaction kernel is also dependent on the $k$-grid density, as
the integrations described in Eqn. \ref{b2} and \ref{b3} are evaluated by a
plane-wave summation in reciprocal space with the help of Fourier transform.
Typically, the band dispersion doesn't change significantly from DFT to GW
bandstructure. Thus one should be able to estimate the $k$-grid density
required to converge a BSE absorption spectra only by observing the change in
the RPA absorption spectra. For the convergence of electron-hole interaction
kernel, we note that in the literature \cite{kammerlander2012speeding} it has
been suggested that $K^{eh}$ varies little w.r.t the $k$-points, as the
single-particle wave functions are quite robust w.r.t $\bf k$. 

Therefore, one can assume that the $k$-mesh required to achieve the convergence
solely from a change in band dispersion (which can be estimated from RPA
calculations) is likely to be sufficient to converge $K^{eh}$ and also the BSE
absorption spectra. In the following paragraph, we will discuss the results
from our calculation, which supports the aforementioned hypothesis. Moreover,
we want to emphasize that by convergent absorption spectra we mean that not
only the positions of the absorption peaks are converged but the oscillator
strengths of these peaks are converged as well so that we obtain a
$\epsilon_2(\omega)$ that doesn't change with the finer sampling of the BZ. 

\begin{figure}[h]
\centering
\includegraphics[width=1.0\linewidth]{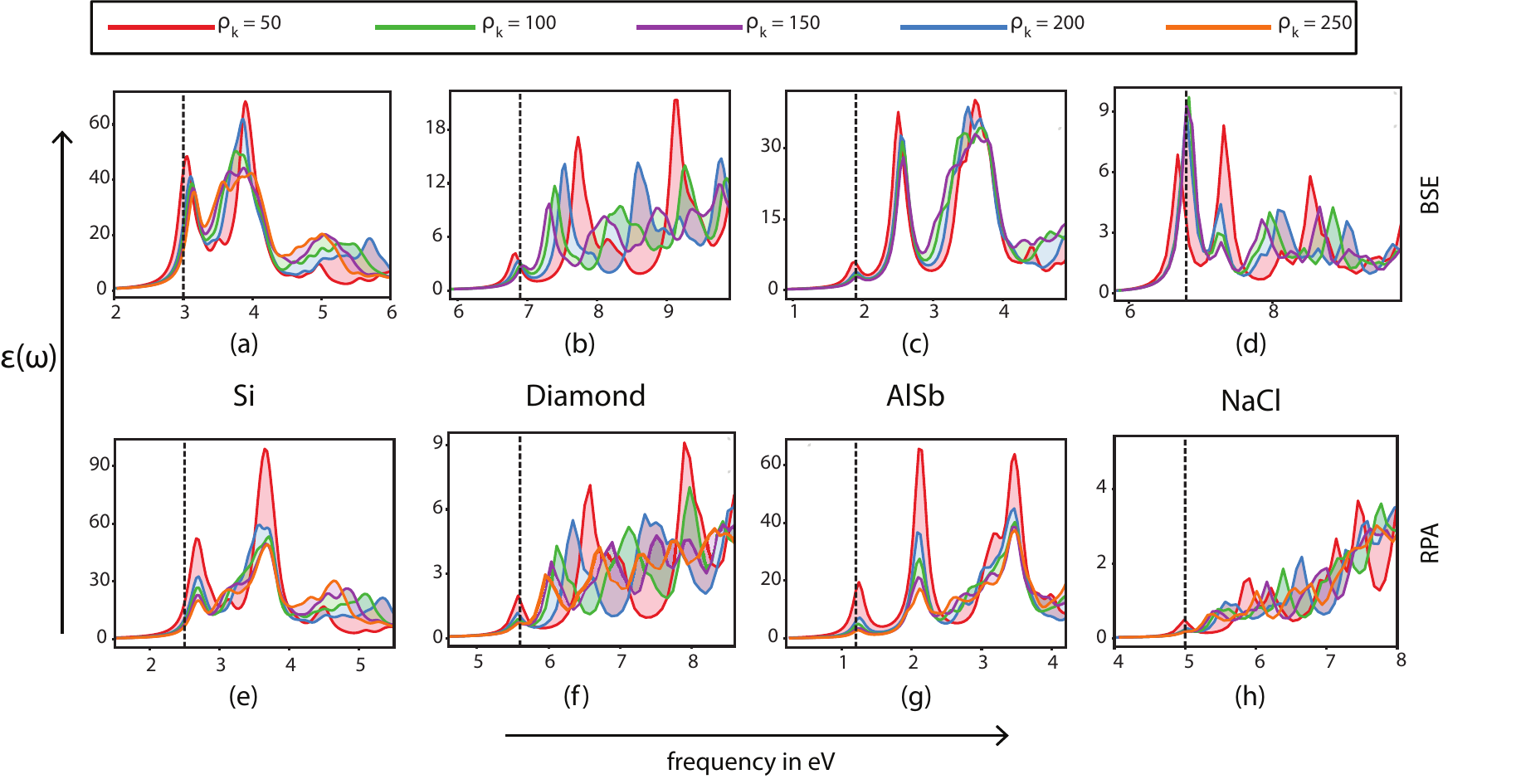}
\caption{Convergence of BSE (top panel) and RPA/DFT (bottom panel) absorption
spectra w.r.t reciprocal density ($\rho_k$) for Si(a,e), Diamond(b,f),
AlSb(c,g) and NaCl(d,h). The optical gap in case of BSE and direct DFT gap has
been shown with a dashed vertical line in each figures.} 
\label{fig2}
\end{figure}

To achieve this, we propose a similarity coefficient, SC, that is a
measure of the convergence of the absorption spectra. We define the similarity coefficient as follows,

\begin{equation}
\mathrm{SC}(\Delta \rho_k, \rho_k) = 1 - \frac {\Delta A (\Delta \rho_k, \rho_k)} {A (\rho_k)}
\label{scdef}
\end{equation}

where, $\Delta A (\Delta \rho_k, \rho_k)$ is the area between two $\epsilon(\omega)$ curves computed with
reciprocal density $\rho_k$ and $\rho_k+\Delta \rho_k$ and $A(\rho_k)$ is
the total area under the $\epsilon(\omega)$ curves computed with reciprocal
density $\rho_k$. Note that $\Delta A$ in Eqn. \ref{scdef} is not simply the difference in area between two curves but quantifies the similarity between two curves by summing up the areas where two curves differ from each other \cite{jekel2019similarity}. This is shown as  the shaded areas in Fig. \ref{fig2}). 

Fig. \ref{fig2} shows the convergence of the absorption spectra of Silicon, Diamond, AlSb, and
NaCl calculated using different reciprocal densities, from both RPA (bottom
panel) and BSE (top panel) calculations. Notably, the absorption spectra from RPA and BSE look drastically different. However, this is expected as the
inclusion of GW corrections increases the bandgaps significantly resulting in a
shift of the entire spectra towards higher frequency. Whereas, the inclusion of
excitonic effects through BSE results in a change in the oscillator strengths
(peak heights) in $\epsilon(\omega)$. In the case of Si, Diamond, and AlSb the
consequences of including excitonic effects are relatively low as the relative
heights of the low energy absorption peaks don't change significantly (Fig.
~\ref{fig2}(a-c) and (e-g)), whereas, in the case of NaCl it shows a very
prominent low energy excitonic peak, almost absent in the RPA spectra (Fig.
~\ref{fig2}(d) and (h)). 

\begin{figure}
\centering
\includegraphics[width=0.8\linewidth]{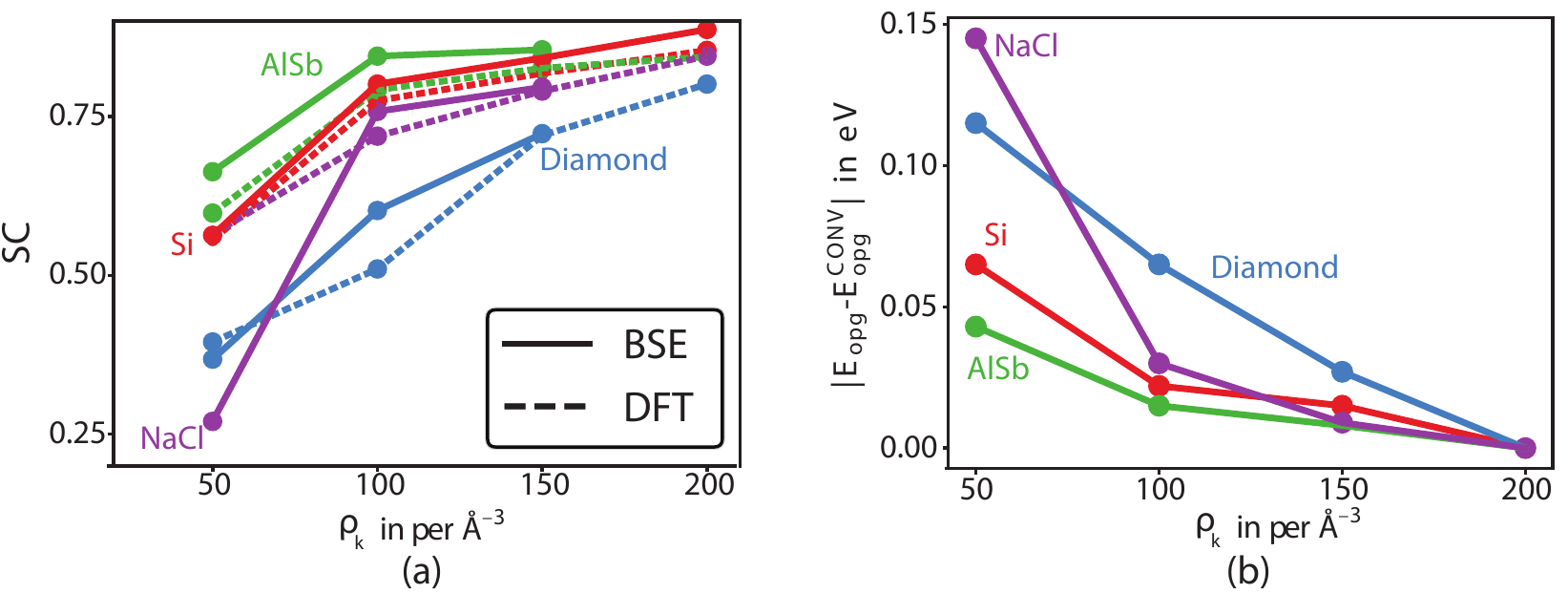}
\caption{Quantifying the convergence of BSE calculation w.r.t reciprocal
density ($\rho_k$) through (a) an area similarity coefficient (SC) for spectra
and (b) optical gap (E$_{opg}$). E$^{CONV}_{opg}$ is the respective converged
value of optical gap for Si (red), Diamond (blue), AlSb (green) and NaCl
(purple) based on our calculation. We have shown that the SC computed from DFT
(RPA) absorption spectra can be used to check the convergence of BSE
calculation w.r.t $\rho_k$.} 
\label{fig3}
\end{figure}

 Fig. \ref{fig3} (a) shows the SC for the spectra of Fig. \ref{fig2} computed with a $\Delta \rho_k$ of 50 per $\mathrm \AA^{-3}$. The SC captures both the shifts in peak positions and oscillator strengths in the absorption spectra resulting from the change in reciprocal density used in
BSE calculations. As one can see from the absorption spectra of Si and AlSb, Fig.
~\ref{fig2}(a) and (c), that the peak positions and their oscillator strengths don't
change too much when $\rho_k$ is $>$ 100 per $\mathrm \AA^{-3}$. This is reflected
by a larger value of SC, $>$ 0.75, for them (Fig. ~\ref{fig3}(a)). Whereas, a
lower SC value is obtained for diamond and especially for NaCl even for $\rho_k$
 is $>$ 100 per $\mathrm \AA^{-3}$. This is mostly due to a large change in peak
positions for diamond and a change in oscillator strengths for NaCl (Fig.
~\ref{fig2}(b) and (d)). Nevertheless, once we look at the convergence of SC
for both BSE and RPA absorption spectra (Fig. ~\ref{fig3}(a)), they look quite
similar. Thus the lower resource and time-intensive RPA can be employed to
estimate the $k$-mesh required to converge the BSE calculations In Fig.
~\ref{fig3} (b) we show the convergence of optical gap, $E_{opg}$, with
the reciprocal density, $\rho_k$, used in BSE calculation. To compare the
convergence for these materials that have very different optical gaps we 
subtracted the converged value of the optical gap, $E^{CONV}_{opg}$, for each of
them to show the variation in the same scale. From Fig. \ref{fig3} (b) it is
clear that the $k$-mesh required to obtain an SC $> 0.75$ is sufficient to
converge the optical gaps within 50 meV for most of the materials. Additionally, Fig.~\ref{fig3}(a) indicates that the convergence of the SC ensures that the entire absorption spectra (in the
desired frequency range) is also converged along with the optical gap.

\section{$py$GWBSE workflow architecture}
\label{workflow}

\begin{figure}
\centering
\includegraphics[width=0.9\linewidth]{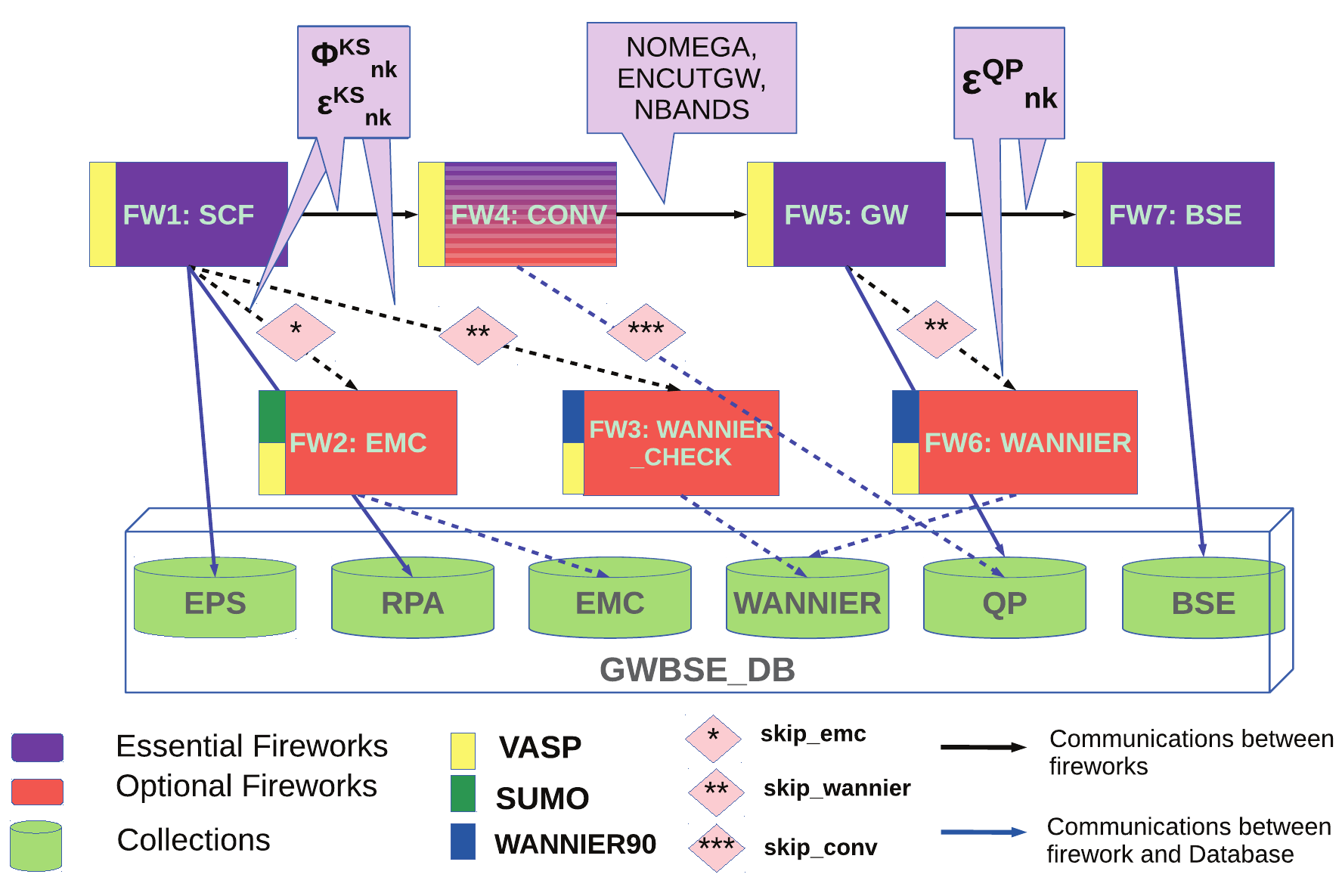}
\caption{All the FWs, both essential (purple rectangles) and optional (orange
rectangles) involved in GW-BSE workflow as implemented in $py$GWBSE code are
shown in the flowchart. The input tags that trigger the optional FWs are shown by pink diamonds. The collections that store input parameters and output
results from different FWs in the GWBSE database are shown by green cylinders.
Essential (solid) and optional (dashed) connections of the FWs with the GWBSE
database (blue) and with other FWs (black) are shown with lines. Crucial
information communicated between fireworks is shown in light purple
callout boxes. Three simulation software $VASP$, $Sumo$ and $Wannier90$ used
by the different FWs. They are shown by yellow, green, and blue side bars, respectively, on the
individual FWs.} \label{fig4}
\end{figure}

Figure~\ref{fig4} shows the workflow architecture of $py$GWBSE. The workflow
consists of seven fireworks (FW). FWs are a set of tasks, called firetasks
(FTs), that together accomplish a specific objective such as DFT structural
relaxation. For example, FW1 named SCF is composed of four FTs (see Figure S3
in supplementary information), which enables a DFT simulation by automatically
generating the input files, running a simulation on a supercomputing resource,
analyzing the simulation output, and storing it in a MongoDB database. Some FWs
are optional in the workflow (shown as orange rectangles in Fig~\ref{fig4})
while others are essential (shown as purple rectangles in the same figure).
Once each of the FWs is completed, the results along with the input parameters
necessary to reproduce the results are saved in a MongoDB database.

FWs and FTs were first introduced in the $Fireworks$ \cite{jain2015Fireworks}
open-source package. They allow us to break down and organize a workflow in a
group of tasks with the correct order of execution for each task and suitable
transfer of information between the tasks. For example, FTs can be simple tasks
such as writing files, copying files from a previous directory, or more complex
tasks such as starting and monitoring a $VASP$ calculation, or parsing specific
information from $VASP$ output files and saving them in a MongoDB database.  

$py$GWBSE creates and stores information about FWs, FTs, and their
interdependencies in MongoDB database collections as JSON objects. These
collections are shown as green cylinders in Fig.~\ref{fig4}. At the time of
workflow execution on a supercomputer, the FTs of individual FWs are executed
in the appropriate order using the JSON objects stored in the MongoDB database.
$py$GWBSE workflow needs a file named `input.yaml' to initiate the workflow. In
`Running the example' section of supplementary information, we have shown an
example `input.yaml' file with a detailed description of all the input tags one
need to specify for creating a workflow that demonstrates the capabilities of
the $py$GWBSE package. 

\subsection{$py$GWBSE fireworks}
\label{fireworks}

As mentioned earlier, we have developed seven FWs in the $py$GWBSE package.
They are shown as rectangular boxes in Fig. \ref{fig4}. Each FW is named as
shown in the boxes along with an `FW' suffix. The FWs are numbered in order of
their execution in the workflow. Three simulation software namely $VASP$,
$Sumo$ and $Wannier90$ are used by $py$GWBSE. The software used for individual
FWs are denoted by left-sided bars on the FW boxes in Fig. \ref{fig4}.

The first FW, ScfFW, is used to obtain self-consistent charge density by
solving the KS equation. It computes KS eigenvalues and wavefunctions, ($\Phi^{KS}_{n\bf k}$ and $\epsilon^{KS}_{n\bf k}$, which are required by all
other FWs. Second FW, EmcFW, performs the effective mass calculation (EMC) via the
$Sumo$ code\cite{ganose2018sumo} using the DFT bandstructure obtained by FW1. The third FW,
Wannier$\_$checkFW is designed for checking the accuracy of wannier
interpolation. Fourth FW, ConvFW performs convergence tests for {\it NBANDS},
$ENCUTGW$, and $NOMEGA$ as discussed in Section \ref{GW methods}. Fifth FW,
GwFW performs GW calculation to obtain QP energies ($\epsilon^{KS}_{n\bf k}$)
as discussed in section \ref{GW methods} (Eqn. \ref{g3} --\ref{g9}). The sixth
FW, WannierFW is designed to perform wannier interpolation to obtain QP
energies along the high-symmetry k-path to produce GW bandstructure. Lastly,
the seventh FW, BseFW solves the Bethe-Salpeter equation to obtain
$\epsilon_2(\omega)$ as described in Section \ref{BSE methods} (Eqn.
\ref{b1}--\ref{b4}).

The optional FWs are triggered by the input tags described in the pink
diamonds. For example, only when $skip\_emc$ tag is set to $False$, the
workflow executes  EmcFW (FW2) to compute the effective masses. Note that
ConvFW (FW4) is labeled as both an essential and optional FW. In the
supplementary information (`Detailed description of $py$GWBSE fireworks'
section) we explain which parts of the FW4 are optional. 

Figure S3 in the supplementary information shows the breakdown of each of the 7
FWs of $py$GWBSE workflow into its constituent FTs. There are 4 categories of
FTs depending on their functionality. The first category can be considered as file
handling FTs (shown as green boxes in Figure S3). These FTs are used to create
or copy files. For example, the $WriteVaspFromIOSet$ FT in FW1, FW2, FW3, and
FW4 are used to write the INCAR, KPOINTS, POSCAR, and POTCAR input files for
the $VASP$ simulations. The second category is that of the simulation FTs (yellow
boxes in Figure S3). These FTs launch an executable to run specific simulations
on a supercomputer. For example, $Run\_Vasp$ is used to run the $VASP$
software. The third category of FTs, communication FTs (shown in light purple
boxes in Figure S3), enables communication between different FTs. For
example, $PasscalcLocs$ in FW1 and FW5 is used to pass the address of the
directory where a parent FW was executed to its children's FWs. The last
category of FTs, the transfer to database FTs, transfers information to
the database. They are shown by gray boxes For example $Eps2db$ in FW1 is used
to read dielectric tensor from $VASP$ output file and save it to the database. 

We use separate MongoDB collections to store the data and metadata, including
inputs and outputs, associated with the FWs and FTs. Fig.~\ref{fig4} shows these various
collections. The group of all these collections is called the GWBSE$\_$DB.
Dielectric tensors, KS eigenvalues, projections of the KS wavefunctions
onto atomic orbitals (for computing projected DOS), RPA dielectric functions
from DFT calculation, effective masses, wannier interpolated bandstructures
from both DFT and GW levels, all the QP energies including those during
convergence, frequency-dependent dielectric function for different light
polarization axis from BSE calculations are some of the key quantities stored
in the collections. 

A more elaborate description of the code's features and the functionality
implementation, especially the convergence of the GW and BSE-related parameters can
be found in the `Detailed description of $py$GWBSE fireworks' section of the
supplementary information. Moreover, the supplementary information includes an
example Jupyter Notebook that shows a step-by-step setup process to create a
workflow and analyze the results obtained from the workflow to determine QP
properties and the BSE absorption spectra of the wurtzite phase of AlN.

\begin{figure}
\centering
\includegraphics[width=1.0\linewidth]{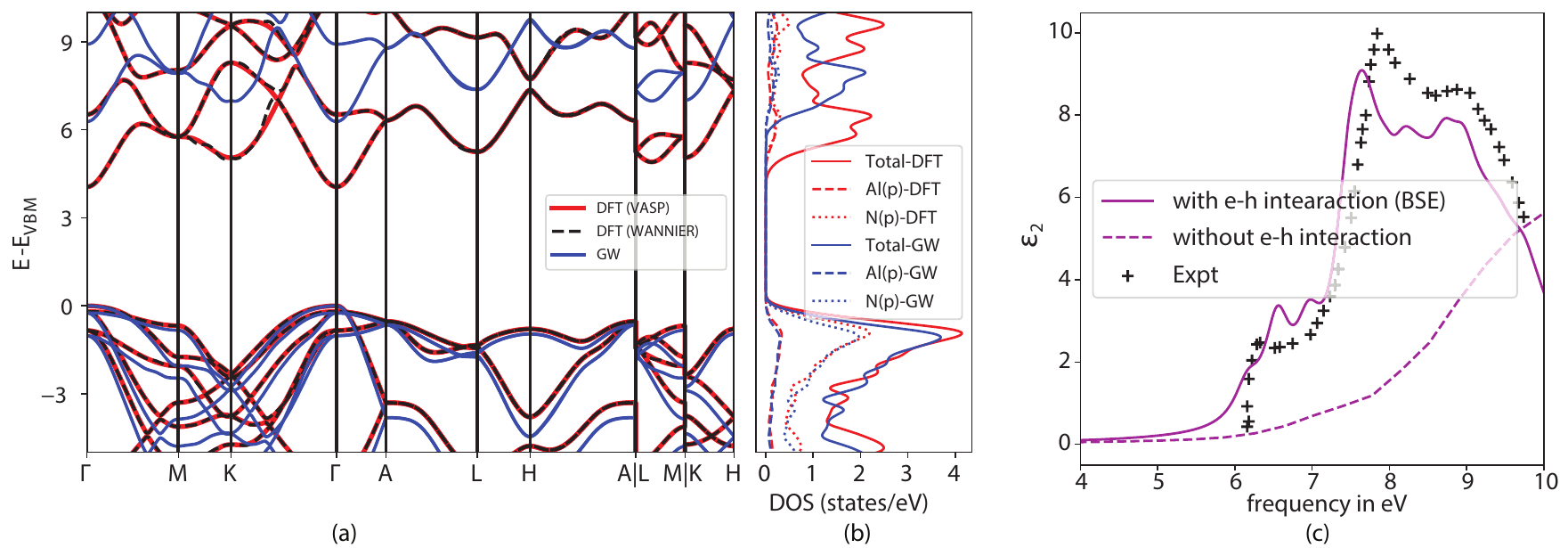}
\caption{(a) Bandstructure of wurtzite-AlN computed using different methods.
DFT bandstructure using direct $VASP$ calculation is shown by red lines and the one obtained from Wannier
interpolation is shown as black dashed lines. The QP bandstructure using partial
self-consistent GW is shown as blue lines. (b) Total and
orbital-resolved DOS for Al (dashed line) and N (dotted line) $p$-states for AlN with
both DFT (red) and GW (blue) calculations. (c) Absorption spectra of
wurtzite-AlN with (solid) and without (dashed) taking electron-hole interaction into
account. The experimental result~\cite{wethkamp1999dielectric} for the same is shown with black `$+$' symbols for comparison.} 
\label{fig5}
\end{figure}

\section{Benchmarking $py$GWBSE for wurtzite AlN}
\label{benchmarking}

Recently, we employed $py$GWBSE to compute the excitonic effects in absorption 
spectra of $\sim$ 50 photocatalysts using the Bethe–Salpeter formalism \cite{biswas2021excitonic}.
In that study, we have compared the QP gap computed using $py$GWBSE for 10
materials of very different chemical compositions with the experimental values 
for the purpose of benchmarking and found excellent agreement. However, in the 
aforementioned study we haven't utilized all the functionalities of $py$GWBSE. 
Therefore, here we demonstrate all the functionalities of the $py$GWBSE workflow 
by applying it to a test case of wurtzite-AlN. In this section, we compare the 
various quantities obtained from the workflow simulations with the experimentally 
measured values reported in the literature. 

We begin by evaluating the quantities obtained from the DFT calculations. To
that end, the dielectric constants of AlN are found to be
$\epsilon_{\infty}^{\perp}$=4.61, $\epsilon_{\infty}^{\parallel}$=4.82 , and
$\epsilon^{avg}_{\infty}$= 4.68, with the $\epsilon^{avg}_{\infty}$ being in
exact agreement with the experimentally measured value of 4.68
\cite{akasaki1967infrared}. 

The electron effective masses are  found to be $m^{\parallel}_e$=0.28$m_0$ and
$m^{\perp}_e$=0.3$m_0$ which fall in the experimentally obtained ranges of
$\sim$ 0.29-0.45$m_0$ \cite{dreyer2013effects}. In case of
hole effective masses we find a  large anisotropy, $m^{\parallel}_h$=0.24$m_0$ and
$m^{\perp}_h$=4.32$m_0$.  The average hole effective mass in AlN was recently
estimated to be $\sim$ 2.7$m_0$, based on experimental measurements of the Mg
acceptor binding energy in Mg-doped AlN epilayers \cite{nam2003mg} which is
very similar to our average computed value of $m^{avg}_h$=2.96. 

Fig. \ref{fig5} (a) shows the bandstructure of AlN computed from DFT using
$VASP$ directly (red solid) and through the use of wannier interpolation (black
dashed). As one can see that, the wannier interpolation is very accurate and
both the bandstructures overlap with each other throughout most of the BZ. Fig.
\ref{fig5} (a) also shows that AlN is a direct gap semiconductor with a DFT gap
of 4.05, which is expectedly underestimated compared to the experimental value
of 6.2 eV \cite{rubio1993quasiparticle} but very close to the value obtained
from DFT calculations in the earlier studies (3.9
eV)\cite{rubio1993quasiparticle}. 

Once we perform the one-shot GW calculation the direct gap increases to 5.59 eV.
Although this is closer to the experimental value it is still not quite
accurate. Previous studies, using an LDA functional as a starting point found a
QP gap of 5.8 eV, which also doesn't agree with the experimental gap. However,
after we perform partial self-consistent GW (scGW) the QP gap becomes 6.28 eV,
resulting in an excellent agreement with the experimental value of 6.2 eV
\cite{rubio1993quasiparticle}. The QP bandstructure with scGW is shown by the blue curve in Fig. \ref{fig5}
(a). 

We use the QP energies and projection of KS wavefunctions or atomic orbitals to
compute the orbital resolved DOS with QP corrections under the assumption that
the KS wavefunction is a good approximation for the QP wavefunction. Fig.
\ref{fig5} (b) shows the orbital-resolved DOS of wurtzite-AlN with Al($p$) and
N($p$) states shown with dashed and dotted lines respectively. We show
orbital-resolved DOS obtained from both DFT and GW calculations for comparison.
Our calculation suggests that the valence band edge of wurtzite-AlN mostly
consists of N($p$) states, whereas the conduction band edge is resulting from
strong hybridization between Al($p$) and N($p$) states, which is consistent
with the findings of previous studies \cite{jiao2011comparison}. 

To show $py$GWBSE's ability to perform BSE calculation and obtain absorption
spectra ($\epsilon_2(\omega)$) that include electron-hole interactions we
perform the GW-BSE calculation for wurtzite-AlN. Fig. \ref{fig5} (c) compares
the absorption spectra ($\epsilon_2(\omega)$) that we obtained from the BSE
calculation, the calculation without electron-hole interaction, and the
experimentally obtained spectra from the literature
\cite{wethkamp1999dielectric}. The light polarization is set to be
perpendicular to the $c$-axis. As we can see from Fig. \ref{fig5} (c) the
absorption spectra calculated without taking electron-hole (e-h) interaction
into account completely misses the features in the 6--10 eV range, visible in
the experimental absorption spectra (shown with `$+$' symbols in Fig.
\ref{fig5} (c). Only when we include the e-h interaction through the BSE
calculation those excitonic features are retrieved. Although, the absorption
spectra obtained from BSE very closely resemble the experimental absorption
spectra we find that the sharp absorption edge at 6.2 eV and two prominent
absorption peaks at 7.85 and 8.95 eV are shifted (by $\sim$ 0.15 eV for the
peaks) to the lower frequencies. We
have used a $\rho_k$ value of 200 per $\mathrm{\AA^{-3}}$
(12$\times$12$\times$7 $k$-grid) with a broadening ($CSHIFT$) \cite{VASP} of
0.2 eV for our AlN BSE calculation, which produces an SC value of 0.89 (see Fig.
S4 in supplementary information for convergence results). In the figures in the SI, we can see that with an SC of $\sim$ 0.75 reasonable amount of information is obtained for the spectra peaks and their positions, however, clearly larger SC's lead to better accuracy, albeit at a higher computational cost and computing time. Thus previous GW-BSE calculations with a finer sampling of the BZ, with randomly distributed 1000 $k$-points, led to even better agreement with the experimental spectra.\cite{bechstedt2005quasiparticle}

\section{Conclusion}
\label{conclusion}

We have developed a Python toolkit, $py$GWBSE, which enables high-throughput
GW-BSE calculations. In this article, we present the underlying theory, the
workflow architecture, the algorithmic implementation, and benchmark
simulations for the $py$GWBSE code. This open-source code (available at ${\it
https://github.com/cmdlab/pyGWBSE}$) enables automated input file generation,
submission to supercomputing platforms, analysis of post-simulation data, and
storage of metadata and data in a MongoDB database. Moreover, $py$GWBSE is
capable of handling multiple convergence parameters associated with the GW-BSE
formalism. To reduce the computational cost associated with obtaining a
converged absorption spectrum from BSE calculations, we present a novel
strategy for computing the similarity coefficient from RPA spectra. We have
shown that this approach ensures convergence of not only the optical gap or
exciton binding energy but the entire absorption spectra in the desired
frequency range. Our openly available code will help to include QP properties
and excitonic effects in future computational material design and discovery
studies in a variety of fields such as power electronics, photovoltaics, and
photocatalysis. The $py$GWBSE  will facilitate high-throughput GW-BSE
simulations enabling the application of large data methods to further explore
our understanding of materials as well as first-principles methods that are
designed for computing excited state properties. 

\section{Acknowledgements}
\label{Acknowledgements}

This work was supported by ULTRA, an Energy Frontier Research Center funded by
the U.S. Department of Energy (DOE), Office of Science, Basic Energy Sciences
(BES), under Award $\#$ DE-SC0021230. In addition, Singh acknowledges support by
the Arizona State University start-up funds. The authors acknowledge the San
Diego Supercomputer Center under the NSF-XSEDE Award No. DMR150006 and the
Research Computing at Arizona State University for providing HPC resources.
This research used resources of the National Energy Research Scientific
Computing Center, a DOE Office of Science User Facility supported by the Office
of Science of the U.S. Department of Energy under Contract No.
DE-AC02-05CH11231. The authors also thank Tara M. Boland, Adway Gupta, Akash
Patel, and Cody Milne for testing of the code and helpful discussions.


\end{document}